\documentclass[twocolumn,superscriptaddress,preprintnumbers,amsmath,10pt,aps,prl,nobibnotes,longbibliography]{revtex4-1} 
\usepackage{amsmath}
\usepackage[ruled,vlined]{algorithm2e}
\usepackage{amsfonts}
\usepackage{bm}
\usepackage{xcolor}
\usepackage{verbatim}
\usepackage{graphicx}
\usepackage{siunitx}
\usepackage{upgreek}

\begin{document}

\title{Terahertz imaging through emissivity control}

\author{Michal Mrnka}
\email{M.Mrnka@exeter.ac.uk}
\author{Harry Penketh}
\author{Ian R.~Hooper}
\author{Sonal Saxena}
\affiliation{Department of Physics and Astronomy, University of Exeter, Exeter, EX4 4QL, UK.}
\author{Nicholas E.~Grant}
\affiliation{School of Engineering, University of Warwick, Coventry, CV4 7AL, UK.}
\author{John D.~Murphy}
\affiliation{School of Engineering, University of Warwick, Coventry, CV4 7AL, UK.}
\author{David B.~Phillips}
\author{Euan Hendry}
\affiliation{Department of Physics and Astronomy, University of Exeter, Exeter, EX4 4QL, UK.}


\begin{abstract}
\noindent{Adoption of terahertz (THz) technologies is hindered by the lack of cost-effective THz sources. Here we demonstrate a fundamentally new way to generate and control THz radiation, via spatio-temporal emissivity modulation. By patterning the optical photoexcitation of a surface-passivated silicon wafer, we locally control the free-electron density, and thereby pattern the wafer’s emissivity in the THz
part of the electromagnetic spectrum. We show how this unconventional source of controllable THz radiation enables a new form of incoherent computational THz imaging. We use it to image various concealed objects, demonstrating this scheme has the penetrating capability of state-of-the-art THz imaging approaches, without the requirement of femto-second pulsed laser sources. Furthermore, the incoherent nature of thermal radiation also ensures the obtained images are free of interference artifacts. Our spatio-temporal emissivity control paves the way towards a new family of long-wavelength structured illumination, imaging and spectroscopy systems.}

\end{abstract}

\maketitle

\noindent {\bf \large Introduction}\\

\noindent{All bodies above absolute zero temperatures are sources of  electromagnetic radiation described by Planck's law~\cite{Planck1901}. The total radiated power is given by the thermodynamic temperature of the object $T$ and its emissivity $\varepsilon$, a material parameter that represents its capacity to emit thermal radiation. Dynamic control of thermal radiation has been a very active research area in infrared spectral bands, with a straightforward concept: if one can find a way to spatially or temporally modulate either the temperature or the emissivity, one can then control the radiated power in space and time~\cite{Coppens2017, Greffet2011, Baranov2019, Picardi2023}. Control of thermal emission has been achieved with temperature modulation~\cite{Wojszvzyk2021, Li2021a}, dynamic modulation of emissivity in bulk~\cite{Xu2019, Xiao2019} and meta-materials~\cite{Inoue2014,Coppens2017} using visible and ultra-violet (UV) light~\cite{Coppens2017}, electrical modulation~\cite{Vassant2013, Brar2015, Park2018}, and using magnetic~\cite{Caratenuto2021} or thermal~\cite{Tang2020} fields. In the infrared (IR) band, this approach has lead to super-Planckian emitters~\cite{Xiao2022}, control of polarisation~\cite{Wojszvzyk2021,Wang2023, Nguyen2023}, control of thermal transport~\cite{Li2021}, new forms of IR camouflage~\cite{Xu2018, Gui2022, Qu2018, Hu2021}, radiative cooling~\cite{Li2018a}, thermography~\cite{Xiao2020} and holography~\cite{Zhou2021}.}

The majority of these approaches are naturally optimised for modulation of mid-infrared thermal radiation, corresponding to peak thermal emission at room temperature for most materials. At the same time, it is appealing to develop thermal modulation approaches for the problematic far-infrared (i.e.\ THz) region of the spectrum ($\sim0.3-3$\,THz) where very few competing technologies exist. Selective generation, control and detection of THz radiation continues to be technologically challenging, with most approaches still relying on the generation and detection of THz radiation by high power or femtosecond lasers~\cite{CastroCamus2021}. This is despite a broad range of prospective imaging and measurement applications lying in wait, stretching from medicine~\cite{Yu2019}, security~\cite{Appleby2007} and quality control~\cite{AfsahHejri2019}, to agriculture~\cite{AfsahHejri2019} and semiconductor industries~\cite{True2021}.

In this paper, we present the first experimental demonstration of spatio-temporal control of thermal emission and detection in the THz frequency band. We utilise illumination by structured visible light to induce spatio-temporal modulation of the non-equilibrium free electron and hole densities in passivated silicon. For rapid modulation of THz emissivity, semiconductors are the natural choice, as free-electron densities can be photo-modulated, while $\sim$THz plasma frequencies typically give rise to large thermal emissivity in the THz range. This brings about local modulation of the emissivity, with a spatial resolution determined by the diffusion length of photocarriers.  An essential feature of our new approach is the surface passivation of the silicon, which dramatically enhances the emissivity photomodulation in the THz spectral range. Using a single detector for the emitted thermal radiation and a sequence of orthogonal illumination patterns, we demonstrate a new incoherent THz imaging method which harnesses this control of thermal emission, by reconstructing THz transmittance images of objects placed in the vicinity of the silicon modulator, without the need for any external THz source. As the photomodulation of thermal emission is maximal in the lower THz band, i.e.\ below 1.5~THz, this approach can be used to image concealed objects undetectable with visible light, and with a resolution unattainable using microwave radiation. \\

\newpage
\noindent {\bf \large Results}\\
\hfill \break
\noindent {\bf Imaging concept and experimental setup}\\
\noindent{A simplified schematic of our  experimental setup is shown in Fig.~1a. A visible beam of wavelength 623~$\upmu$m, generated by an LED, is spatially and temporally patterned by a digital micromirror device (DMD) with small individually controllable mirrors. The DMD chip is imaged onto a surface-passivated, high-resistivity silicon wafer. The visible light beam with photon energy above the band gap of silicon generates free electron-hole pairs, which locally modulates the emissivity. Right side of Fig.~1a shows how the contrast between the illuminated and unilluminated areas is created -- each small illuminated region behaves like a local Lambertian source with the maximum power radiated in the direction normal to the wafer. 

An object to be imaged is placed close to the surface-passivated wafer. We mount a thin, thermally insulating layer (of thickness $\approx$200\,$\upmu$m) in between the object and the wafer to prevent conductive heating of the object. The thermal radiation emitted by the wafer is transmitted through the object and collected by a large 90$^\circ$ offset aluminium paraboloid mirror with an effective focal length 101.6\,mm and an equally large diameter. The numerical aperture (NA) of the system on the object side of the mirror is NA = 0.2 resulting in a diffraction limited spot size of 0.75\,mm at 1\,THz.  The THz radiation transmitted through the object is collected by the mirror and coupled to an f/2 Winston cone with 25\,mm diameter through two low-pass optical filters to reject radiation above 1.5\,THz. The aperture of the Winston cone is located at a plane conjugate to the plane of the wafer. Inside a He cryostat, the Winston cone is followed by an InSb bolometer, with detection frequency bandwidth 0.06-1.5\,THz and response time of 350\,ns. The signal from the detector is amplified by a low-noise electronic amplifier. A 3\,kHz electronic low-pass filter is used to reduce the instantaneous bandwidth to better suit our imaging rates while suppressing the noise level. In the experiments, we apply a small tilt of ~15\,$^\circ$ to the wafer to redirect the specular reflection toward a room temperature absorber instead of the detector itself, which prevents unwanted artefacts (see Supplementary information, section 8).  The temporal voltage signal is digitised with an analogue-to-digital converter, and the signal collected for a sequence of different patterned visible beams enabling the image to be reconstructed (see Methods and Fig.~1b for example image).

\begin{figure}[ht!]
\centering
\includegraphics[width=1\linewidth]{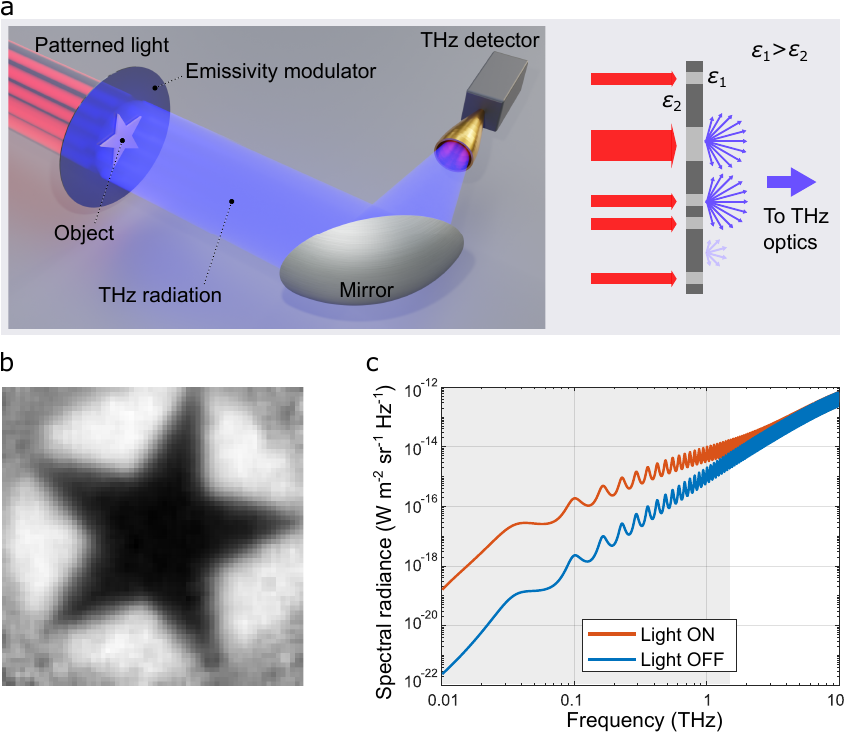}
\caption{\textbf{Principle and experimental setup.} \textbf{(a)} A spatially-patterned visible light beam illuminates a silicon wafer from one side. The structured terahertz beam generated by the modulated emissivity  of the wafer $\varepsilon$ (on the right) interacts with the object and is collected by the large offset parabolic mirror to be focused onto a single-pixel detector. \textbf{(b)} A measured image of a conductive object, low brightness corresponds to low THz transmission. \textbf{(c)} Simulated spectral radiance of the passivated silicon wafer with 675\,$\upmu$m thickness and 0.36\,ms effective carrier lifetime  at 290\,K temperature. The red and blue curves correspond to the radiances with and without applying a photoexcitation light intensity of 200\,W/m$^2$. The difference between these curves gives rise to the photomodulation efficiency.  The gray area corresponds to the spectral range of our detector. }
\label{fig:1}
\end{figure}

\hfill \break
\noindent {\bf Emissivity modulation}\\
\noindent{
 The key to achieving THz imaging through emissivity control is the design of our THz photo-modulator.  In the experiment, the patterned visible beam illuminates a 675\,$\upmu$m thick, high-resistivity, undoped silicon wafer engineered with a ZnO/Al$_2$O$_3$ surface passivation layer stack. This passivation reduces the surface recombination velocity of generated free carriers which therefore increases the effective charge carrier lifetime $\tau_\mathrm{eff}$ from $\approx0.022$\,ms (unpassivated) to $\approx0.36$\,ms (passivated) \cite{Hooper2019,Hooper2022}. We note, that without the Al$_2$O$_3$ capping layer, ZnO does not provide sufficient passivation to have any impact on the modulation efficiency. In this case the Al$_2$O$_3$ layer provides a source of hydrogen~\cite{Loo2019} which enables improved surface passivation and an optimal effective lifetime of $\approx0.36$\,ms. The level of passivation given by ZnO/Al$_2$O$_3$ was thus appropriate here as it enables intermediate lifetimes to be reached between unpassivated and excellent surface passivation given by Al$_2$O$_3$ alone.The cross section of the wafer is given in the supplementary (Fig.~4) and a detailed description of its fabrication in the Methods section.} 

When light illuminates the wafer, photo generated electron-hole pairs modify the permittivity of the material. This in turn renders the modified silicon wafer an efficient absorber/emitter in the millimetre and lower terahertz bands via an increase in the plasma frequency of the electrons and holes (see supplementary material section 4). The high effective carrier lifetime in the passivated wafer provides a substantial increase in emissivity at terahertz frequencies under a modest illumination intensity ($\approx$100s of Watts per m$^2$) -- see Fig.~1c, where the spectral radiance $B(\nu, T)$ is calculated according to supplementary section 2 and $\nu$ is the temporal frequency. The spectral radiance expected for a silicon wafer with a 0.36\,ms  effective carrier lifetime is plotted for two cases: with and without photoexcitation intensity of 200\,W/m$^2$, corresponding to electron densities of $6.38\cdot10^{20}$\,m$^{-3}$ and $1.45\cdot10^{16}$\,m$^{-3}$, respectively. The emission peaks in Fig.~1c correspond to Fabry-Pérot resonances, and are calculated for emission normal to the wafer. Note that, since we have a collection angle of $\pm 11.5\,^\circ$ (NA = 0.2), and these peaks shift with angle, they will be diminished in the experiment. As the emissivity modulation depth is proportional to the difference between the radiance of the photoexcited and non-photoexcited silicon, the useful spectral range of the emission approximately coincides with the spectral range of the detector (gray area in Fig.~1c).

In order to further boost the THz thermal emission, we increase the thermodynamic temperature of the wafer to $120\,^{\circ}$\,C  heated by convection using a heat gun facing the illuminated face of the wafer. We use thermocouple devices on both sides to monitor the temperature gradient across the device. We observe that the emission signal increases with increasing temperature almost linearly, as expected by the Rayleigh-Jeans law (see Fig.~2). Kirchhoff's law explains why heating is required to increase signals (see supplementary section 4).

\begin{figure}[ht!]
\centering
\includegraphics[width=0.75\linewidth]{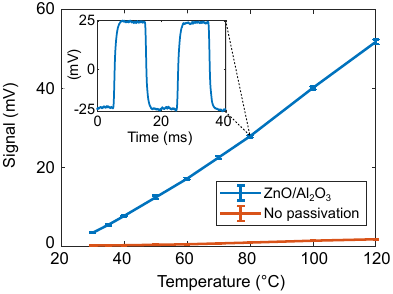}
\caption{\textbf{Thermal emission enhancement via surface passivation.} Measured signal level as a function of wafer temperature for the ZnO/Al$_2$O$_3$ passivated wafer with a room temperature carrier lifetime $\tau$ = 0.36\,ms and for the unpassivated wafer with $\tau$~=~0.022\,ms.  The on/off photoexcitation is applied with intensity of $300$\,W/m$^2$. Inset shows the modulated signal sampled in time for the passivated wafer.  The measured signal level corresponds to the amplitude of the spectral component at the modulation frequency.}
\label{fig:2}
\end{figure}


\newpage
\noindent {\bf Imaging}\\
\noindent{We image several different objects under a photoexcitation intensity of 200\,W/m$^2$. Figures~3(a-c) show images of a copper etched Siemens star deposited on a thin plastic film and placed $\approx$200\,$\upmu$m away from the wafer, chosen to test the imaging resolution of our technique. The THz image is normalised to an image taken without the sample present under the same conditions. Dark regions of the image correspond to areas with low THz transmission -- in this case shielded by copper. The size of the full field-of-view  of the system is 27 by 27\,mm$^2$. Here we set the pixel size to 0.42x0.42\,mm$^2$ (controlled by how many DMD micro-mirrors are binned into a single super-pixel. The minimum discernible feature in the image is about ~1.2 mm (see the red arrows in Fig.~3b), limited by free-charge diffusion in the wafer.

One of the main advantages of terahertz radiation is its ability to penetrate through certain visibly opaque materials such as paper, plastics and ceramics. Here, we demonstrate our system is sensitive enough to take advantage of this capability, by imaging two hidden objects shown in Figs~3(d-g): a razor blade in a paper envelope and a commercial radio frequency identification (RFID) tag sandwiched between a plastic and a paper layer. For noise suppression, we apply a simple 2D sliding average filter with a mask size of 3x3 pixels -- every pixel in the final images is thus a mean value of 3x3 neighbouring pixels from the original image. The circular copper wiring with outside diameter of 22\,mm, and the chip inside of the wiring, are both opaque for the terahertz radiation, while the plastic cover of the tag is mostly transparent at these frequencies. The two narrow leads connecting the chip and the wiring can be faintly recognized in the upper left-hand corner of the tag.

\begin{figure*}[ht!]
\centering
\includegraphics[width=0.65\linewidth]{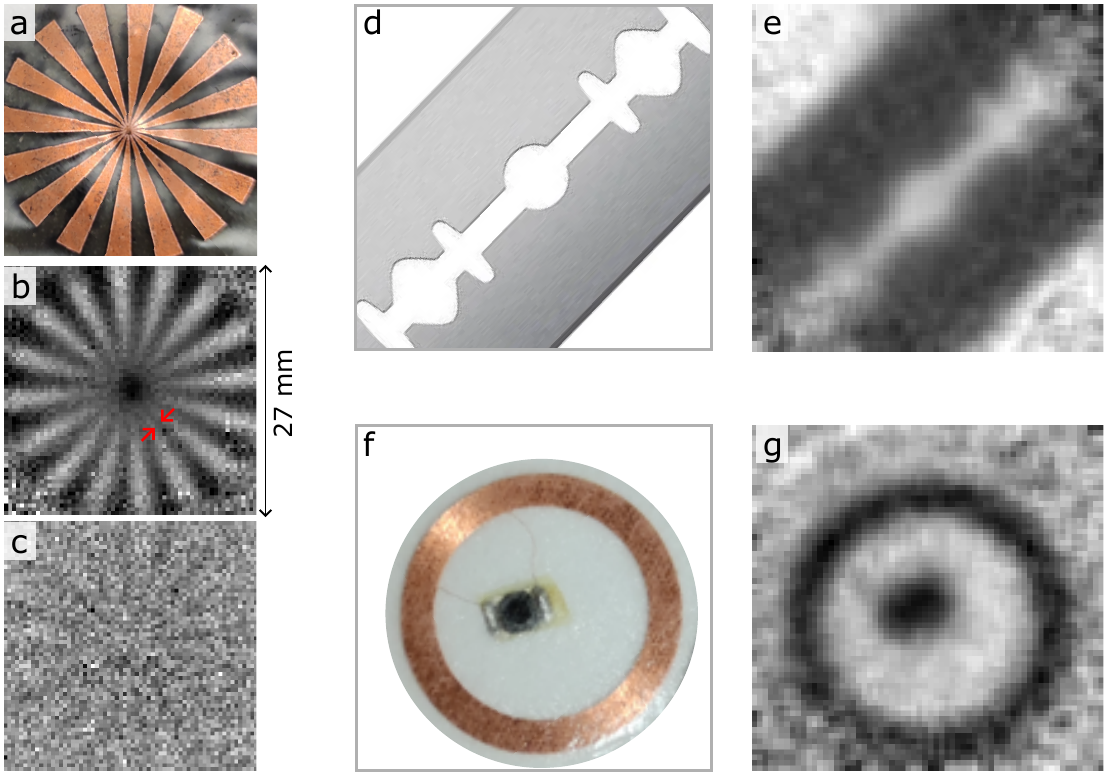}
\caption{ \textbf{Imaging with the emissivity modulator.} (\textbf{a-c}) Images of a Siemens star resolution target - Optical (\textbf{a}) and terahertz (\textbf{b}) images, the latter reconstructed with our ZnO/Al$_2$O$_3$ passivated wafer. For comparison, image (\textbf{c}) is measured using an unpassivated high-resistivity silicon wafer. Red arrows in (\textbf{b}) highlight the smallest resolvable feature, estimated to be 1.2\,mm. (\textbf{d-g}) Optical images of a razor blade (\textbf{d}) and an uncovered radio frequency identification tag (\textbf{f}). \textbf{e} and \textbf{g} show our THz images of these objects concealed inside a paper envelope.}

\label{fig:3}
\end{figure*}

\hfill \break
\noindent {\bf \large Discussion}\\
One of the key considerations in this work is the optimisation of the emissivity photomodulator. There are competing constraints that lead to a trade-of between different aspects of device performance. To achieve optimal image resolution, a short carrier lifetime should be prioritised. This is because image resolution is currently determined by carrier diffusion length  $L = \sqrt{D\tau_\mathrm{eff}}$, where $D$ is the diffusion coefficient of charge carriers. By engineering a wafer with a longer effective lifetime we trade a shorter integration time for a lower image resolution (see Supplementary information, section 4).  We can see the resolution in Fig.~3b (which is estimated as 1.2\,mm) is notably lower compared to the diffraction limit (0.75\,mm at 1~THz) and agrees with the expected diffusion length of electrons of ~1.1\,mm for a carrier lifetime of 0.36\,ms.

However, there is simultaneously a stringent trade-off between imaging time and image resolution in our approach. A significantly shorter carrier lifetime results in a much lower modulation of emissivity for the same optical power. For comparison, in Fig.~3c we show an image with an unpassivated 22\,$\upmu$s effective lifetime wafer, taken under the same conditions as Fig.~3b. Furthermore, the fact that the signal increases with the increasing carrier lifetime (see Fig.~2 and supplementary information, section 4) demonstrates that it is the modulated emissivity and not temperature that gives rise to the increased thermal emission on photoexcitation. 

Finally, a carrier lifetime of $\tau_\mathrm{eff}\approx 0.36$\,ms allows us to implement a 10\,ms pattern projection time for each optical pattern, resulting in an imaging time of approximately 41 seconds for 64x64 pixels. However, in our proof-of-principle experiments, in order to reduce image noise, averaging over much longer times is currently required: the images in Figs.~3b/3c and 3e/3g were integrated for 176 and 44 mins, respectively. We note that these long integration times are a current limitation of our detector, with signal to noise determined by its noise equivalent power NEP $\approx0.13\,\mathrm{nW}/\mathrm{Hz}^{1/2}$ (see supplementary section 9). A modest 3-fold improvement in sensitivity would result in almost an order-of-mangitude improvement in integration times. Likewise, a 3-fold increase in the optical illumination intensity would yield a similar result. Furthermore, if priors about the object under inspection are available, compressive sensing can also be used to reduce the number of patterns that need to be projected, thus enhancing the imaging rate~\cite{Stantchev2017}. We therefore emphasize that there is significant scope for future improvement in imaging times with this approach.

Finally, we note that Caratenuto et al.~\cite{Caratenuto2021} recently published simulation results that suggest that modulation of thermal THz radiation is feasible using a high magnetic field applied to an InSb based metamaterial to tune a THz emissivity resonance. However, it is not clear how an approach relying on high magnetic fields can be extended to spatially varying control or how fast modulation may be achieved.\\

\hfill \break
\noindent{\bf \large Conclusions}\\
In summary, we have demonstrated a dynamic, all-optical, spatial emissivity modulator at terahertz frequencies based on photomodulation of surface-passivated silicon. We have also proposed and experimentally validated an imaging method that inherently relies on such a modulator, requiring no external THz source. Thanks to the relatively long charge carrier lifetime of our surface-passivated Si wafer, the emissivity modulation required only a modest intensity patterned pump beam (200\,W/m$^2$) - provided by a standard LED source and a digital micromirror device. For imaging, the emissivity modulator was used as a source of spatially-patterned, thermally radiated (i.e.\ broadband, incoherent) terahertz radiation that illuminated samples placed in close proximity. By detecting the signal in a ghost imaging setup~\cite{Padgett2017}, we demonstrated a convenient imaging technique based on a low-cost terahertz source and a single-pixel detector.

We believe our imaging technique has potential to compete with standard THz imaging approaches, overcoming the current dependence on femtosecond lasers for generation and detection in this frequency band. 
Furthermore, by combining thermal emission modulation with microantenna arrays patterned onto the wafer, additional control may be achievable, including over spectral content, directionality~\cite{Liu2011,Greffet2002, Shin2019, Zhang2019, Fenollosa2019} and polarisation state~\cite{Wojszvzyk2021,Wang2023, Nguyen2023} of the thermally emitted terahertz radiation. These possibilities suggest our thermal modulation approach will find a range of applications beyond imaging.\\

\noindent {\bf \large Methods}\\
\noindent {\bf Sample fabrication}\\
\noindent{The silicon based modulators were fabricated from 675\,$\upmu$m thick, FZ $>$5000 $\Omega$cm n-type wafers with a diameter of 125\,mm. Prior to dielectric/metal-oxide deposition, the to-be passivated silicon wafer underwent a thorough surface preparation procedure in order to minimise organic and metal particulate contamination on its surfaces. This involved dipping the wafer in a HF (2\%) solution to remove any native oxide film, followed by an RCA 1 clean (H$_2$O, H$_2$O$_2$ (30\%), NH$_4$OH (30\%) in a ratio 5:1:1) at $\sim$75$^\circ$C for 10 minutes, a dip in HF (2\%) to remove the chemical oxide, an RCA 2 clean (H$_2$O, H$_2$O$_2$ (30\%), HCl (37\%) in a ratio 5:1:1) at ~75$^\circ$C for 10 minutes followed by a final HF (2\%) dip to remove the chemical oxide. The wafers were pulled dry from the final HF solution (no rinse) and immediately loaded into the load lock of the Veeco Fiji G2 atomic layer deposition (ALD) system. 30\,nm (200 cycles) of zinc oxide (ZnO) was then deposited at 200$^\circ$C by thermal ALD using diethylzinc and water as the precursors. This process was repeated on both surfaces of the silicon wafer. Following the ZnO depositions, 20\,nm (160 cycles) of aluminium oxide (Al$_2$O$_3$) was deposited (at 200$^\circ$C) on top of the ZnO layer using an O$_2$ plasma source and a trimethylaluminium precursor. This process was also repeated on both surfaces of the silicon wafer. Following the ALD depositions, the wafers underwent an activation anneal at 440$^\circ$C for varying durations to achieve the desired level of passivation and thus effective lifetime (see SI for details). }\\

\noindent {\bf Single-pixel imaging}\\
The imaging technique presented in the paper is based on a single-pixel architecture with structured illumination~\cite{Edgar2018}, sometimes referred to as computational ghost imaging~\cite{Padgett2017}. An unknown object is sequentially illuminated by a set of spatially-structured, mutually-orthogonal binary patterns. The light transmitted through the sample is detected as a temporal signal with a single-pixel  detector. This provides a significant simplification of the detection with respect to standard array detectors. Knowing the illumination patterns (which are held in vectorised form along the columns of matrix $\mathrm{\mathbf{P}}$) and the corresponding detected temporal signals from the bucket detector 
 (held in column vector $\mathrm{\mathbf{s}}$), which carries the information on the interaction of the patterned THz beam with the object, one can reconstruct an image of the unknown object (expressed in vectorised form as $\mathrm{\mathbf{o}}$) by simple matrix multiplication:
\begin{equation}
    \mathrm{\mathbf{o}} =  \mathrm{\mathbf{P}}^{-1} \mathrm{\mathbf{s}},
\end{equation}
provided the inverse of the patterns, $\mathrm{\mathbf{P}}^{-1}$, is known. In this work, 2D reshaped rows of Hadamard matrices are used as the individual structured patterns as they are mutually orthogonal and each projection measurement collects light from at least half of the total number of the image pixels (i.e.\ half of the field of view of the scene) which results in a much higher signal to noise ratio compared to e.g.\ raster scanning. \\

\noindent {\bf \large Acknowledgments}\\
The authors acknowledge financial support from the Engineering and Physical Sciences Research Council (EP/S036466/1, EP/W003341/1, EP/R004781/1, EP/V047914/1 and EP/S036261/1).
DBP thanks the European Research Council (804626), and the Royal Academy of Engineering for financial support.\\

\noindent {\bf \large Author contributions}\\
M.M. and E.H. devised the concept of the modulator and the experiment. M.M. carried out the analytical modelling, data acquisition, post-processing and wrote the first draft of the manuscript. D.B.P. and H.P. provided support on computational imaging aspects of the project and assisted with the design of the data acquisition software. M.M.,E.H., H.P., D.B.P, I.R.H. and S.S. contributed to the interpretation of the experimental data. I.R.H. assisted with the measurements of the effective lifetimes. N.E.G. and J.D.M. designed and fabricated the emissivity modulator. All authors edited the manuscript.\\

\noindent {\bf \large Competing interests}\\
 The authors declare that there are no conflicts of interest related to this article.\\

 \noindent {\bf \large Data availability}\\
 The data that support the findings of this study are available from the corresponding author upon request\\

 \noindent {\bf \large Additional information}\\
 Correspondence and requests for information should be addressed to M.M.\\


\newpage
\noindent {\bf \large References}\\

\bibliography{main}

\onecolumngrid
\setcounter{equation}{0}
\setcounter{figure}{0}
\vspace{50cm}
\noindent{\centering{\LARGE Supplementary Information - Terahertz imaging through emissivity control}\par}
\vspace{5mm}


\author{Michal Mrnka}
\email{M.Mrnka@exeter.ac.uk}
\author{Harry Penketh}
\author{Ian R.~Hooper}
\author{Sonal Saxena}
\affiliation{Department of Physics and Astronomy, University of Exeter, Exeter, EX4 4QL, UK.}
\author{Nicholas E.~Grant}
\affiliation{School of Engineering, University of Warwick, Coventry, CV4 7AL, UK.}
\author{John D.~Murphy}
\affiliation{School of Engineering, University of Warwick, Coventry, CV4 7AL, UK.}
\author{David B.~Phillips}
\author{Euan Hendry}
\affiliation{Department of Physics and Astronomy, University of Exeter, Exeter, EX4 4QL, UK.}


\noindent{This supplementary document provides further information on the theory, various trade-offs and design of the terahertz emissivity modulator and the imaging method described in the paper. The experimental setup and methods used to characterise the wafers are explained here. We also discuss specific imaging artifacts.}\\

\section{1. Experimental setup}
\noindent{The detailed experimental setup is shown in Fig.~1. The emissivity modulator (undoped Si wafer) is covered by a 0.5\,mm thick quartz glass on the heated side for protection. A heat gun (DIAFIELD 2000\,W) with variable temperature setting heats up the wafer via convection. A Thorlabs 4.8\,W, 623 ±5\,nm LED is used as a source for the photoexcitattion pattern projections and a digital micromirror device (DMD)  Vialux, V7001 with 1024 x 768 pixels generates the modulation patterns that are projected onto the surface of the wafer through the protective glass. During the experiments, only a central,  square part of the DMD with 768 x 768 pixels is used. A 60\,mm focal length lens with 1" diameter refocuses the patterns generated by the DMD with the right magnification to achieve the 27x27\,mm$^2$ field of view. The wafer is tilted by ~15$^\circ$ with respect to the optical axis so that the specular reflection from the wafer is redirected towards a room temperature absorber, with another absorber placed on the optical axis behind the wafer. The object is placed on the non-heated side of the modulator. A 101.6\,mm diameter ($f$ = 101.6\,mm) offset paraboloid mirror 250\,mm away from the wafer is used to collect the THz radiation and focuses it onto a  cryogenic, 4.2\,K InSb hot electron bolometer with magnetic tuning (QMC QFI XBI) sensitive in the frequency range 0.06-1.5\,THz. The DMD triggers the data acquisition from the detector and the acquisition and reconstruction of images is performed in Labview environment.}
\begin{figure}[ht!]
\centering
\includegraphics[width=0.58\linewidth]{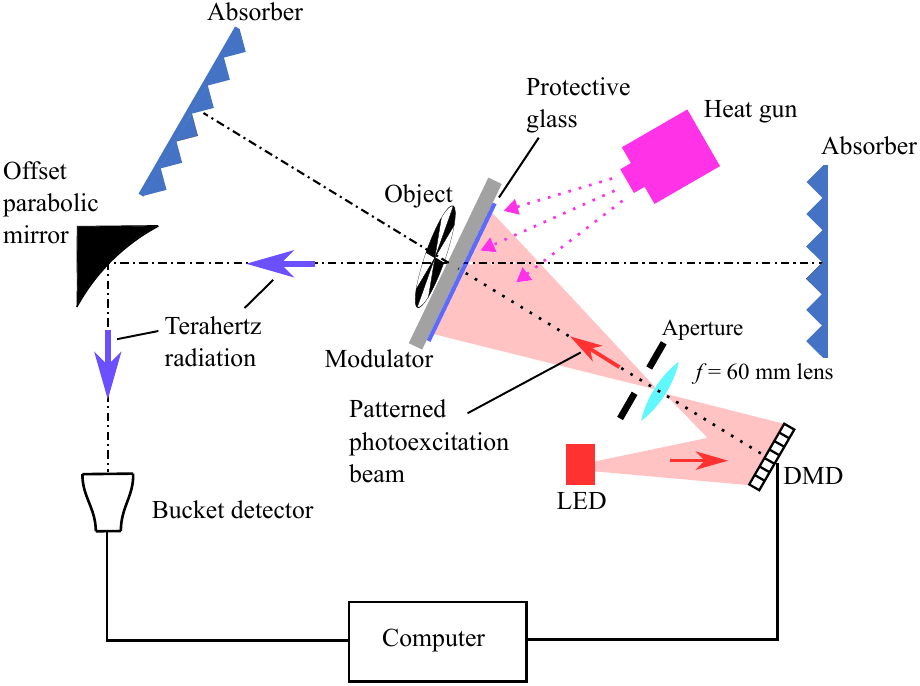}
\caption{Experimental setup.}
\end{figure}

\section{\bf 2. Emissivity modulation}
\noindent{Photomodulation of a heated, surface-passivated, monocrystalline silicon wafer lies at the heart of our imaging method. As such, the properties of intrinsic silicon as well as the way the terahertz and visible light interact with the wafer directly influence the imaging quality that can be achieved. We employ the wafer as a planar terahertz source whose emitted power distribution across its area is modulated by spatially photomodulating the emissivity of the wafer using a digital micromirror device (DMD).}\\

We rely on Kirchhoff's law of thermal radiation \cite{Kirchhoff1860} to explain the method and to estimate its performance. According to Kirchhoff's law of thermal radiation the emissivity of an object in thermal equilibrium with its surrounding equals its absorptivity. This means that the emissivity of the wafer is modulated at the same rate/level as the absorption of the wafer (given by the DMD). In other words, any good terahertz absorption-based transmission modulator \cite{Hooper2019} is suitable as a source for our method. If needed, the Kirchhoff's law can be generalised for non-equilibrium bodies as shown in \cite{Greffet2018}, however, in our work we rely on the approximate original version of the law. 

We can tie together the transmission $\mathcal{T}(\theta)$, reflection $\mathcal{R}(\theta)$ and emissivity $\varepsilon (\theta)$ of the wafer as functions of angle $\theta$ as follows:

\begin{equation}
    \mathcal{T}(\theta) + \mathcal{R}(\theta) + \varepsilon (\theta) = 1.
\end{equation}

The transmission through and reflection from the wafer can be calculated by a transfer matrix method \cite{hecht}, or from a dielectric slab equation \cite{Born1999} in the case the passivation layers are not considered. Assuming region 1 and region 3 (see Fig.~2) have the same material properties:

\begin{equation}
        \mathcal{T} = t  t^* = \left|\frac{t_{12} t_{23} \exp{(\mathrm{i} \beta)}}{1+r_{12} r_{23} \exp{(2 \mathrm{i}} \beta)} \right|^2,
\end{equation}
\begin{equation}
        \mathcal{R} = |r|^2 = \left|\frac{r_{12} + r_{23} \exp{(2\mathrm{i} \beta)}}{1+r_{12} r_{23} \exp{(2 \mathrm{i}} \beta)}\right|^2,
\label{eq:}
\end{equation}
with $t$, $r$ being the transmission and reflection coefficients and
\begin{equation}
\beta = \frac{2\pi}{\lambda} d \cos{\theta_2} = \frac{\omega}{c} d \cos{\theta_2},
\end{equation}
where $r_{12}$, $r_{23}$, $t_{12}$ ,$t_{23}$ are the reflection and transmission coefficients at the interfaces between regions , $\lambda$ is the free space wavelength, $\theta_2$ is the refraction angle in medium with $n_2$ given by Snell's law $\theta_2 = \arcsin(n_1/n_2)\sin\theta$ and $\omega$ is the angular frequency.

Once the transmission and reflection of the wafer are known for given photoexcitaton intensity $I$ and frequency $\nu$, we can calculate the approximate emissivity  from  $\varepsilon(I,\nu) = 1-\mathcal{T}(I,\nu) - \mathcal{R}(I,\nu) $ while disregarding any scattering contributions. The results for the emissivity difference between the photoexcited and unilluminated silicon wafer corresponding to $I$=300 W/m$^2$ and $I$=0\,W/m$^2$ for three different charge carrier lifetimes ranging from 0.03\,ms to 1\,ms are presented in Fig.~3.

\begin{figure}[ht!]
\centering
\includegraphics[width=0.35\linewidth]{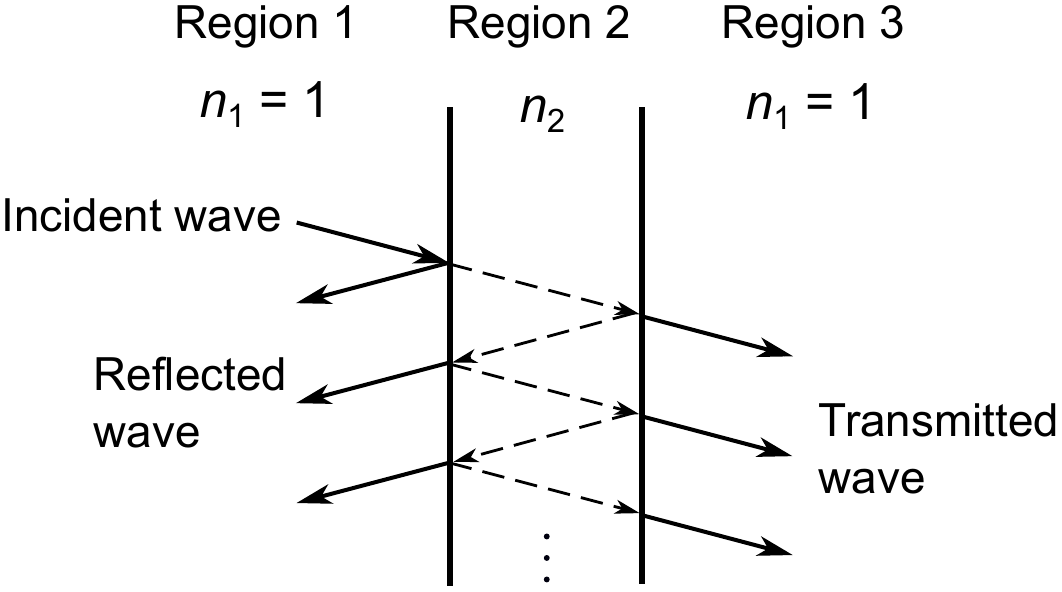}
\caption{Fabry-Pérot cavity formed by a slab of material with refractive index $n_2$ surrounded by materials with $n_1$ and $n_3$.}
\label{fig:fp_cavity}
\end{figure}

\noindent {\bf Emissivity modelling}\\
The changes in emissivity upon photoexcitation can be modelled using Drude's model for relative permittivity $\epsilon$ encompassing the interaction of the semiconductor with light:
\begin{equation}
    \epsilon = \epsilon_0 - \frac{\omega_\mathrm{pe}}{\omega \left( \omega + \mathrm{i}\gamma_\mathrm{e} \right)}   -   \frac{\omega_\mathrm{ph}}{\omega \left( \omega + \mathrm{i}\gamma_\mathrm{h} \right)},
\end{equation}
where $\omega_\mathrm{pe}$ and $\omega_\mathrm{ph}$ are plasma frequencies of electrons and holes, $\epsilon_0$ = 11.7 + 0.003i is the background permittivity of silicon for $\nu \rightarrow \infty$, $\omega = 2 \pi \nu$ is the angular frequency, $\omega_\mathrm{p(e,h)} = \sqrt{(N_\mathrm{0(e,h)} + \Delta n)e^2 / (\epsilon_0 m_\mathrm{(e,h)})  }$, with $N_\mathrm{0(e,h)}$ being the intrinsic carrier concentration at room temperature, $\Delta n = G \tau_\mathrm{eff}$ is the excess carrier density generated by the photo excitation process \cite{Hooper2019}, $\tau_\mathrm{eff}$ is the effective carrier lifetime, $G = T I / (h f d)$ is the generation rate of electron-hole pairs and depends on the intensity $I$ and frequency $f$ of the visible photoexcitation beam and on the thickness $d$ and the transmission through the wafer $T$, $e $ is the elementary charge, $\epsilon_0$ is the vacuum permittivity, $m_\mathrm{(e,h)}$ are the effective masses of the carriers, $\gamma_\mathrm{(e,h)}$ are the scattering rates of electrons  $\gamma_\mathrm{e} = e/m_\mathrm{e}\mu_\mathrm{e}$ and holes $\gamma_\mathrm{h} = e/m_\mathrm{h}\mu_\mathrm{h}$ with $\mu_\mathrm{(e,h)}$ being their mobilities. Many of the parameters given above are temperature dependent, most notably the intrinsic carrier densities, mobilities, effective masses and lifetimes and we analyse the influence of each in more detail in section 4 below. However, in the range of temperatures reported in this paper, it is the electron/hole-phonon scattering that dominates, and the impact of the temperature increase on the achievable permittivity and thus emissivity contrast is relatively small.\\

\begin{figure}[ht!]
\centering
\includegraphics[width=0.5\linewidth]{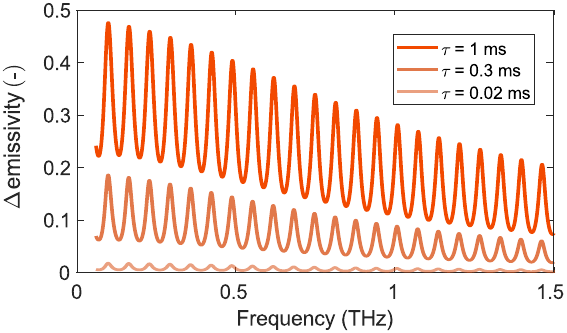}
\caption{Change in emissivity upon photoexcitation for $\theta = 0^{\circ}$ as a function of frequency for increasing carrier lifetimes of an intrinsic Silicon wafer with 675\,$\upmu$m thickness under illumination intensity 300\,W/m$^2$.}
\label{fig:emissivity_modelling}
\end{figure}

\noindent {\bf Thermal radiation}\\
As the emissivity of the wafer is spatially modulated by the visible illumination, so is the spectral radiance $B(\nu, T)$ emanating from the various  regions of the silicon wafer according to 
\begin{equation}
    B(\nu, T) = \varepsilon(\nu) B_\mathrm{bb}(\nu, T),
\end{equation}
where $B_\mathrm{bb}(\nu, T)$ is the spectral radiance, sometimes called brightness, of a black body given by Planck's law \cite{Planck1901}:
\begin{equation}
   B_\mathrm{bb}(\nu, T) =  \frac{2h \nu^3}{c^2} \left( \frac{1}{\mathrm{e}^{\frac{h\nu}{k_\mathrm{B}T}} - 1}  \right),
\end{equation}
with a unit of W\,m$^{-2}$\,sr$^{-1}$\,Hz$^{-1}$ which corresponds to the power thermally radiated by a surface with a unit area per unit solid angle and per unit frequency band. Here $h$ and $k_\mathrm{B}$ are Planck's and Boltzman's constants, $c$ is the speed of light in vacuum, $\nu$ is the temporal frequency and $T$ is the thermodynamic temperature of the black body. It follows that the total radiated power can be determined by integrating $B$ over the surface area, frequency and solid angle. Figure~1c (main article) illustrates how the spectral radiance of a thermal source changes when its emissivity is boosted. This demonstrates how the emissivity modulation  may be preferred to a temperature modulation if a modulated radiant flux is to be achieved, especially if a rapid spatial and temporal modulation is required.
Simple modelling shows that by modifying the lifetime, the silicon becomes an efficient emissivity modulator with the average emissivity contrast $\Delta\varepsilon$ increased by a factor of 22 (from 0.014 to 0.32) at 0.2\,THz and by a factor of 28 (from 0.006 to 0.175) at 1\,THz as the effective lifetime increases from 0.02\,ms to 1\,ms (see Supplementary, Fig.~3)

\section{\bf 3. Surface passivation stack}

In order to determine the absorption and thus emissivity of the passivated wafer (see section 2), we need to first determine the reflection and transmission according to eq.~1. In order to do so, we measure the unpassivated and passivated wafers in transmission geometry under normal incidence in a 60\,GHz quasi-optical setup (for details on the measurement setup see section 5 of the supplementary). First, we measure the transmission through the unpassivated wafer $\mathcal{T}_0$ at normal incidence . Then, we measure the transmission through the passivated wafer $\mathcal{T}_1$. We then tune the conductivity $\sigma$ of ZnO layers ($\epsilon_i = \sigma/(\epsilon_0\omega)$, $\epsilon_0 = 8.854\cdot10^{-12}$\,F/m) in the transfer matrix model to fit the results of the model to the experimental data. We fit the measured transmission change between the unpassivated and passivated wafer $\Delta \mathcal{T}^{\mathrm{measured}}=\mathcal{T}_0^{\mathrm{measured}} - \mathcal{T}_1^{\mathrm{measured}} = 0.349$ to the change predicted by the model $\Delta \mathcal{T}^{\mathrm{model}}=\mathcal{T}_0^{\mathrm{model}} - \mathcal{T}_1^{\mathrm{model}}$ - at the measurement frequency 60\,GHz.

The transfer matrix model includes a simplified 5 layer stack [20\,nm Al$_2$O$_3$, 30\,nm ZnO,  675\,$\upmu$m Si, 30\,nm ZnO, 20\,nm Al$_2$O$_3$]. For modelling Al$_2$O$_3$ we used the values of complex relative permittivity $\epsilon_r = 9.4 + 0.0029$i from \cite{Khalid2008}. A bulk ZnO has a low conductivity ($\approx 10^{-7}$\,S/cm) and a dielectric constant ($\approx \epsilon_\mathrm{r}$ = 8) \cite{Kaur2021}. Unfortunately, when deposited via atomic layer deposition, ZnO's conductivity is significantly increased by incorporated hydrogen impurities  to values roughly about 200\,S/cm \cite{Beh2017}. In the transfer matrix model, the conductivity of the ZnO layers is optimised to determine the conductivity of ZnO in our experiments to be about 290\,S/cm, under the assumption the dielectric constant stays constant in the frequency range of interest. 

Once the conductivity of the ZnO layers is known, we can use it to estimate how the Drude model predicted emissivities in Fig.~3 can be modified by our ZnO/Al$_2$O$_3$  passivation stack- see Fig.~5.

\begin{figure}[h!]
\centering
\includegraphics[width=0.30\linewidth]{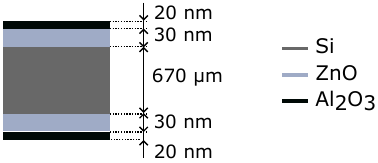}
\caption{The passivation stack designed to increase the emissivity contrast of the high resistivity silicon with the thicknessess of individual layers.}
\label{fig:passivation_stack}
\end{figure}

\begin{figure}[h!]
\centering
\includegraphics[width=1\linewidth]{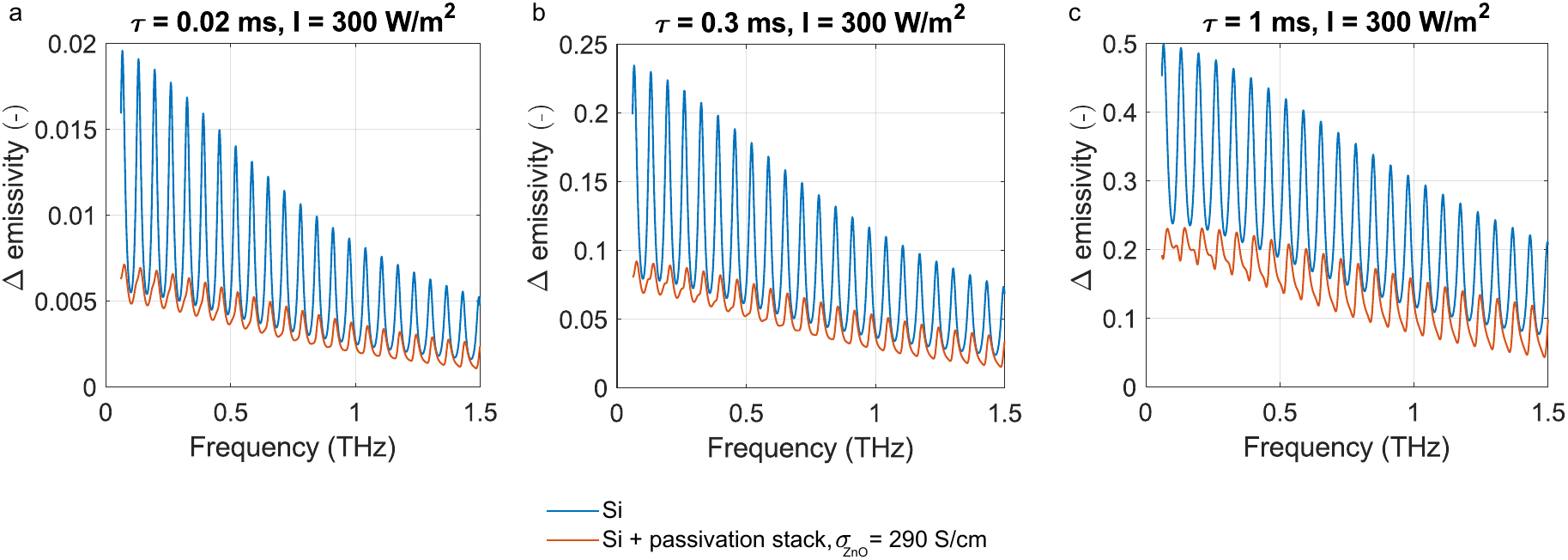}
\caption{Impact of ZnO/Al$_2$O$_3$ passivation on achievable emissivity modulation (difference in emissivity between illuminated and unilluminated silicon) compared to an ideal case (blue curves), where the passivation layers don't influence the emitted terahertz light. The plots correspond to a silicon wafer with increasing effective carrier lifetimes, 0.02\,ms (a), 0.3\,ms (b) and finally 1\,ms (c). }
\end{figure}

\newpage
\section{\bf 4. Effects of wafer heating}
\noindent{In order to improve the signal to noise levels in our study, we increase the temperature of the wafer, so that the changes in emissivity correspond to larger radiant flux modulation. For that we heat up the wafer via convection with a heat gun to an operation temperature of about 120\,$^\circ$C measured with a contact thermocouple  (UNI-T UT320D Mini Contact Type Thermometer). \\}

\begin{figure}[h!]
\centering
\includegraphics[width=0.5\linewidth]{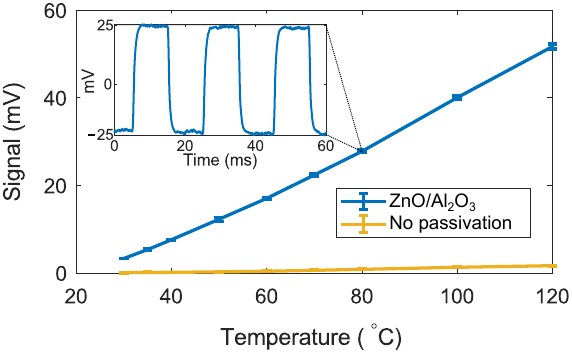}
\caption{Measured signal level as a function of wafer temperature for a passivated wafer with a room temperature lifetime $\tau$=0.36\,ms and for an unpassivated wafer with $\tau$=0.022\,ms. Inset corresponds to a modulated signal sampled in time for a passivated wafer.}
\end{figure}

We apply an on-off modulation with 50\,Hz modulation frequency to the full field of view (FoV) of the detector in the wafer plane and we record the signal level as a function of time (see Fig.~6 inset) and temperature (Fig.~6). As the signal from the detector suffered from a large low-frequency distortion we determined the signal level from the amplitude of the spectral component corresponding to the 50\,Hz repetition rate of the modulation signal and not from measuring the peak-to-peak value of the time domain signal. The detected signal, proportional to the power incident onto the detector, increases with increasing temperature almost linearly, which agrees well with the Rayleigh-Jeans law $B_\mathrm{bb} = 2 k_\mathrm{B} T \nu^2/c^2 $ \, being a good approximation to Planck's law (eq.~7) at low frequencies where $h \nu << k_\mathrm{B} T$. We observe a dramatic increase in the detected signal as the effective lifetime of the carriers increases from 0.02\,ms for the unpassivated to 0.36\,ms for the passivated wafer.

Many of the parameters in the silicon model are temperature dependent, most notably the intrinsic carrier densities, mobilities, effective masses and lifetimes. In the range of temperatures reported in this paper, it is the electron/hole-phonon scattering that dominates, and the impact of the temperature increase on the achievable permittivity and thus emissivity contrast is relatively small. As the emitting medium (predominantly the free-electron cloud) is thick relative to the wavelength, it will support a series of Fabry-Pérot resonances, resulting in the emissivity being a strong function of temporal frequency.

We show here how the intrinsic concentration $n_\mathrm{i}$ and carrier mobility $\mu$ depend on temperature $T$. These two quantities directly influence the achievable emissivity contrast and resolution through diffusion, respectively.

We can model the changes in the intrinsic carrier density of electrons $n_\mathrm{i}$ and holes $p_\mathrm{i}$ with temperature $T$ by \cite{Kittel2005}:
\begin{equation}
        n_\mathrm{i} = p_\mathrm{i} = 2 \left( \frac{k_\mathrm{B} T}{2\pi\hbar^2} \right) ^{3/2}
        \left( m_\mathrm{e} m_\mathrm{h} \right) ^{3/4} \exp{ \left( -E_\mathrm{g} / 2 k_\mathrm{B} T \right)},
\end{equation}
where $k_\mathrm{B} = 1.380649 $×$ 10^{-23} $\,m$^2 $\,kg $ $\,s$^{-2} $\,K$^{-1}$ is the Boltzmann constant, $\hbar =$  1.05457 × 10$^{-34}$\,J\,s is the reduced Planck constant, $m_\mathrm{e}$, $m_\mathrm{h}$ are the effective masses of electrons and holes and $E_\mathrm{g}$ = 1.11\, eV \cite{Kittel2005} is the band gap of silicon at 300\,K. Here, we assume $m_\mathrm{e}$ = 0.26\,$m_\mathrm{0}$, $m_\mathrm{h}$ = 0.38\,$m_\mathrm{0}$ \cite{Kannegulla2015} with $m_\mathrm{0}$ = 9.3109\,x\,10$^{-31}$\,kg being the free-electron mass. The carrier densities of electrons and holes are equal as we consider the case of an intrinsic silicon wafer. 

An alternative model which more accurately fits the experimental data in the temperature range 200\,K $< T < 500\,$K can also be employed \cite{Wasserrab1977}:
\begin{equation}
        n_\mathrm{i} (T) = 6.1 \cdot 10^{19} (T/300)^{2.52} \exp{ \left(-6721/T \right)}.
\end{equation}
Fig.~7 gives the predictions according to eq. 8 and 9. 

The band gap of intrinsic silicon slightly shrinks with temperature, see \cite{Bludau1974}, and it decreases by $\approx 0.03$\,eV as the temperature increases from 300\,K to 400\,K \cite{Thurmond1975}. On the other hand, the conductivity effective mass of electrons and holes \cite{Riffe2002} increases by about 0.0055\,$m_\mathrm{0}$ for electrons and 0.0053\,$m_\mathrm{0}$ for holes in the same temperature range. These correspond to relative changes of about 2\,\% and 1.5\,\% for electrons and holes respectively.\\

The mobilities of electrons and holes decrease as the temperature increases mainly due to phonon scattering \cite{Arora1982}
\begin{equation}
        \mu_\mathrm{e} = 88 \left( T/300 \right)^{-0.57} + \frac{7.4 \cdot 10^8 \, T^{-2.33}}{ 1 + 0.88 \left[  \frac{N_\mathrm{d}}{1.26\cdot10^{17} \left(T/300\right)^{2.4}}   \right]  \left(  T/300 \right)^{-0.146} } \quad \mathrm{cm}^2/\mathrm{Vs},
\end{equation}

\begin{equation}
        \mu_\mathrm{h} = 54.3 \left( T/300 \right)^{-0.57} + \frac{1.36 \cdot 10^8 \, T^{-2.33}}{ 1 + 0.88 \left[  \frac{N_\mathrm{a}}{2.35\cdot10^{17} \left(T/300\right)^{2.4}} \right] \left(  T/300 \right)^{-0.146} } \quad \mathrm{cm}^2/\mathrm{Vs}.
\end{equation}
where $N_\mathrm{d}$ and $N_\mathrm{a}$ correspond to doping concentration. For small doping levels, the mobility becomes insensitive to the doping level, which is the case of the intrinsic silicon.\\

Another temperature dependent parameter directly responsible for the imaging quality is the carrier diffusion length  $L = \sqrt{D\tau_\mathrm{eff}}$, where $D = \mu_\mathrm{e,h} k_\mathrm{B} T /e$ is the diffusion coefficient. Fig.~8 demonstrates how the diffusion coefficient/length and the carrier mobility change as a function of temperature for both electrons and holes.

\begin{figure}[ht!]
\centering
\includegraphics[width=0.8\linewidth]{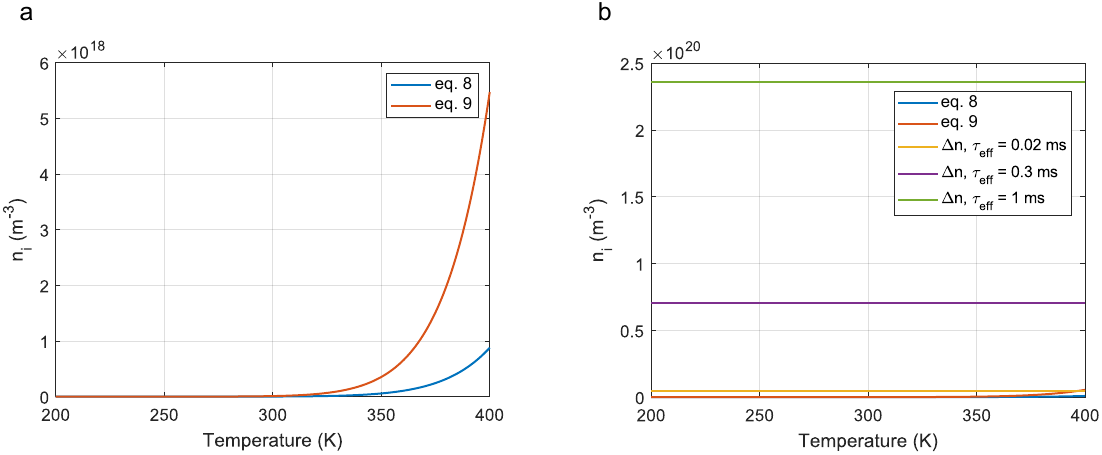}
\caption{(a) Intrinsic carrier concentration dependence on the thermodynamic temperature according to eq. 8 and 9. (b) Comparison of the intrinsic concentrations with the excess carrier densities under 300\,W/m$^2$ illumination intensity.}
\end{figure}

\begin{figure}[ht!]
\centering
\includegraphics[width=0.9\linewidth]{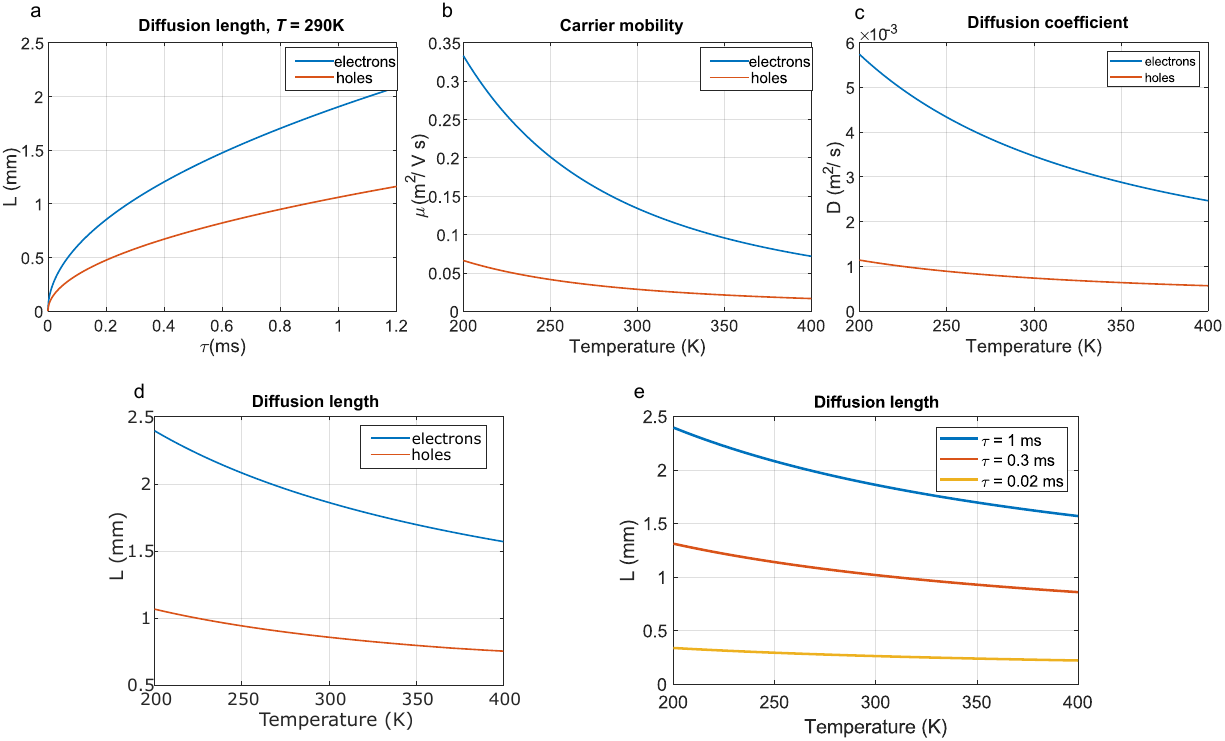}
\caption{(a) Diffusion length as a function of effective carrier lifetime for electrons and holes. (b-c) Carrier mobility and diffusion coefficient as functions of temperature. (d) Diffusion length of electrons and holes as a function of temperature for effective carrier lifetime $\tau_\mathrm{eff}$=1\,ms. (e) Diffusion length of photoexcited electrons as a function of temperature for 3 different effective lifetimes.}
\end{figure}

\noindent {\bf Note on the elevated temperature requirement}\\

Kirchhoff's law can explain why heating of the wafer is necessary and why the change in emissivity of a room temperature wafer is not sufficient to generate any observable signal, even though one might intuitively assume that Planck's law allows it. At room temperature, under a thermal equilibrium condition, the radiation absorbed by the wafer from the thermal emission of the environment is balanced out by the radiation emitted from the wafer. It follows that, in the event of emissivity increase, the additional radiation thermally emitted from the wafer supplants the radiation that would have been transmitted through/reflected from the wafer, but is now absorbed as the absorptance increases equally with emissivity (see eq.~1). It is only when the wafer's temperature is elevated with respect to the environment that the modulation in the emitted signal becomes measurable.

In this regard, it is also important to control what the detector ``sees'' in reflection from the wafer/sample as the reflectivity of the wafer is modulated at the same rate with its emissivity. In section 8 of this supplementary, we demonstrate an experimental artefact, where, at normal incidence, the detector ``sees'' its own reflection, which makes it possible to image with the wafer at room temperature. We deal with this by adding a small tilt ~15$^\circ$ to the wafer and redirecting the specular reflection into an absorbing material.

\noindent {\bf Note on modelling assumptions}\\

Throughout the paper we have relied on the original formulation of Kirchhoff's law, which in turn relies on a thermal equilibrium condition of an isothermal body. In the experiments, we recorded a slight temperature gradient across the depth of the wafers, 170\,mK/$\upmu$m, with temperature of the front surface facing the heat source 413\,K and the back surface facing the sample 393\,K corresponding to average temperature difference of 20\,K across the thickness of 675\,$\upmu$m (wafer) + 500\,$\upmu$m (protective quartz glass) = 1170$\upmu$m. This fact renders the standard Kirchhoff's law an approximation, and a local form of the law \cite{Greffet2018} should be applied to rigorously model the physics of the material. In addition, due to the relatively long switching times described in this study we have disregarded any effects caused by hot carrier dynamics as the hot carrier dynamics are relevant on very short, tens of picoseconds time scales \cite{Xiao2019}, and is thus not accounted for in our models where the average measurement time is on the order of milliseconds. As our focus is on the imaging, rather than on the accurate modelling of the processes in the material we rely on the standard form of Kirchhoff's law in the paper to estimate the emissivity modulation in Fig.~3 (and Fig.~1c of the main paper). We note that to improve the accuracy of the model, the local formulation is preferred, but if used it would not bring any qualitative difference to the outcomes of our paper.

In estimating the emissivity in Fig.~3 we assumed the relative permittivity given by the Drude model does not depend on the depth in the wafer. This is a valid assumption as the relatively long diffusion length in the passivated wafers at 393\,K was simulated to be 1550\,$\upmu$m for the passivated wafer which makes it larger than the wafer's thickness (675\,$\upmu$m). 

\section{\bf 5. Measuring effective charge carrier lifetime }
\noindent{ We measure the effective charge carrier lifetime of the wafers in a quasi-optical 60\,GHz setup with 2 microwave horn antennas 1 m apart facing each other and the silicon wafer placed mid-way between them. The wafer is tilted with respect to both the microwave beam (15$^\circ$) and the photoexcitation beam (15$^\circ$). One of the antennas is connected to a single port of a vector network analyser (Anritsu Vectorstar MS4647B) operating in a CW mode with frequency 60 GHz and the second antenna is connected to a a Sage Millimeter SFD-503753-15SF-P1 waveguide detector. The signal from the detector is then sampled in time by an oscilloscope while the wafer undergoes periodic TTL photomodulation from a visible source (a collimated 4.8W SOLIS-623C LED from Thorlabs with an output wavelength of 623 nm).}

Figure 9 shows the  transmission signal through the 675\,$\upmu$m thick wafers used in the experiments. We measure the signals for 2 wafers - an unpassivated, reference sample, and a surface-passivated wafer used in our experiments to attain higher and lower effective lifetimes. We repeat the experiment at two different temperatures of the wafers at room temperature T = 293\,K and at elevated temperature T = 393\,K. An industrial heater (Diafield 2000W Heat Gun) is used to heat the wafers via convection and their temperauture is measured with a thermocouple. In the plots, at t = 0\,s time, the photoexcitation source is turned off and the effective lifetime can be approximately determined from the relaxation time where the signal level increases to $\Delta T(1-1/e)$ of its original value before t = 0\,s and  $\Delta T$ corresponds to the transmission change upon photoexcitation. The plots are accompanied by curves corresponding to exponential fits to the experimental data. The unpassivated wafer is modulated with optical intensity of 500\,W/m$^2$, whereas the passivated wafer with intensity of 50\,W/m$^2$ and thus the effective carrier lifetime corresponds to the rise-time of the fitted exponentials. For the unpassivated reference wafer, we obtain $\tau$= 22\,$\upmu$s and 360\,$\upmu$s in the case of the ZnO/Al$_2$O$_3$ passivated wafer. Fig.~9 a-b show how the values of rise time increase when the temperature is increased by 100\,K. On the other hand, Fig.~9 c-d show how the rise time and modulation depth change with increasing photoexciting illumination intensity from 50\,W/m$^2$ to 500\,W/m$^2$ for the two temperatures 293\,K (room temperature) and 393\,K. It is important to notice that the modulation depth in the case of the heated wafer is only slightly higher compared to the room temperature wafer despite the substantial increase in the rise time. 

\begin{figure}[ht!]
\centering
\includegraphics[width=0.85\linewidth]{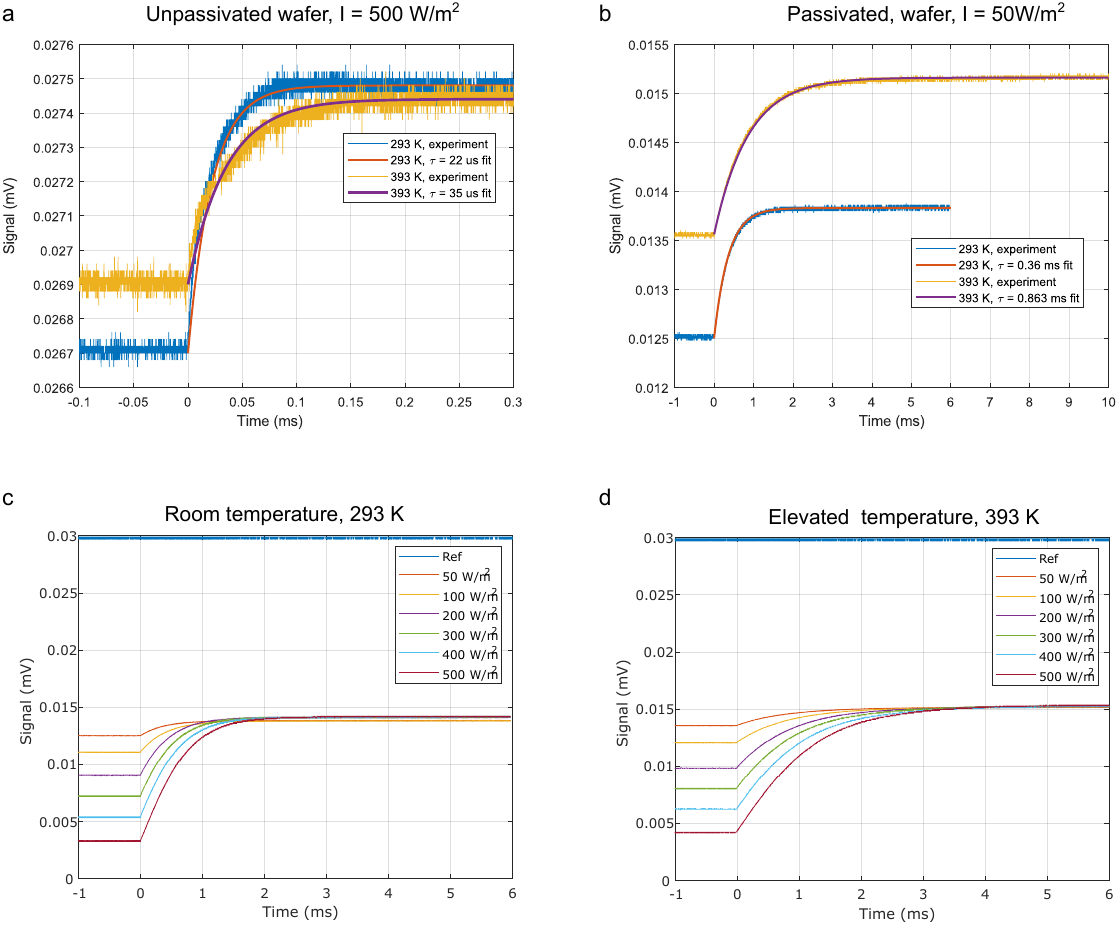}
\caption{Measured switch-off times for an unpassivated 675\,$\upmu$m thick silicon wafer (a) and a ZnO/Al$_2$O$_3$ passivated wafer (b) with the same thickness obtained under 500\,W/m$^2$ and 50\,W/m$^2$ illumination, respectively. Change in the switch-off time upon heating the wafers by 100\,K is also shown. Measured switch-off times for room temperature (c) and heated wafer (d) under variable illumination intensities. }
\end{figure}

\newpage
\noindent {\bf Note on the temperature dependency of the effective charge carrier lifetime }\\

When the temperature of the silicon modulator is either increased or decreased, the effective lifetime of the minority carrier also varies, thereby having a direct impact on the emissivity. There are three ways in which the effective lifetime can increase with temperature (i) by reducing Auger and radiative recombination \cite{Altermatt1997, Altermatt2006}, (ii) by reducing Shockley-Read-Hall recombination in the silicon bulk due to grown in defects and impurities \cite{Rein2002} and (iii) by reducing surface recombination \cite{Eberle2019}. Whilst we cannot determine which recombination mechanism is dominant in our case, we note that this temperature behaviour is expected and thus not an anomaly of the measurement procedure or the design of the modulator.

\section{\bf 6. Imaging of selected targets}
\noindent{Fig.~3 of the main paper includes images of an envelope concealed razor blade and an RFID tag that were filtered to suppress the noise in the raw images. Here, in Fig.~10b we present the original raw image without any post-processing correction and for comparison Fig.~10c shows the results after mean filtering (identical to Fig.~3e of the main paper). Every pixel in the filtered image is calculated as a mean value from 3x3 pixel neighborhood around the corresponding pixel.\\}

Fig.~10d and e present results of a razor blade imaging without any concealment without and with the 3x3 mask mean filtering. One can directly see the increase in the signal-to-noise ratio in the unfiltered image compared to the concealed case in Fig.~10 b. 

Fig.~11b then shows the raw reconstructed image of the RFID tag with 11a and 11b showing the object without concealment in visible range and the final filtered image 11c.

In addition, Fig.~12a-c show the photograph of the copper star object from Fig.~1b of the main paper, the raw reconstructed THz image and a filtered image obtained using the mean filter (identical to Fig.~1b).

\begin{figure}[ht!]
\centering
\includegraphics[width=0.65\linewidth]{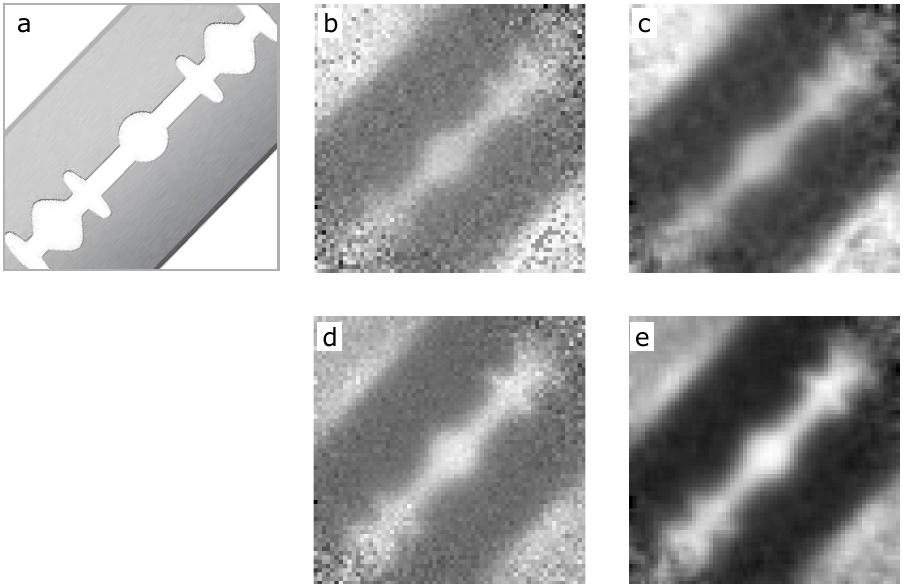}
\caption{(a) Photograph of a razor blade target. (b) 64x64 pixel, terahertz image of the razor blade concealed in a paper envelope without filtering. (c) Mean filtered image. (d) Raw image of the razor without the envelope and after mean filtering (e). }
\end{figure}

\begin{figure}[ht!]
\centering
\includegraphics[width=0.65\linewidth]{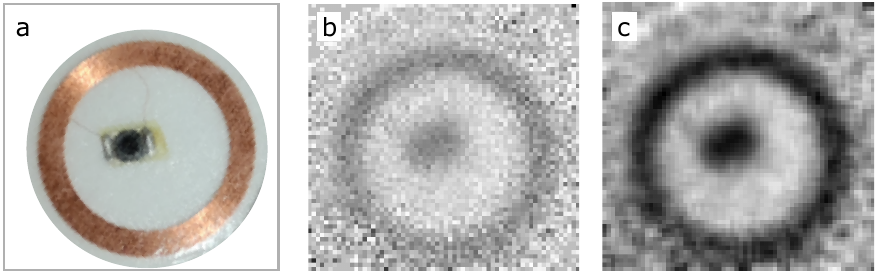}
\caption{(a) Photograph of an RFID tag. (b) 64x64 pixel, terahertz image of the RFID tag concealed in its plastic cover. (c) Filtered image with 3x3 mean filtering mask.}
\end{figure}

\begin{figure}[ht!]
\centering
\includegraphics[width=0.65\linewidth]{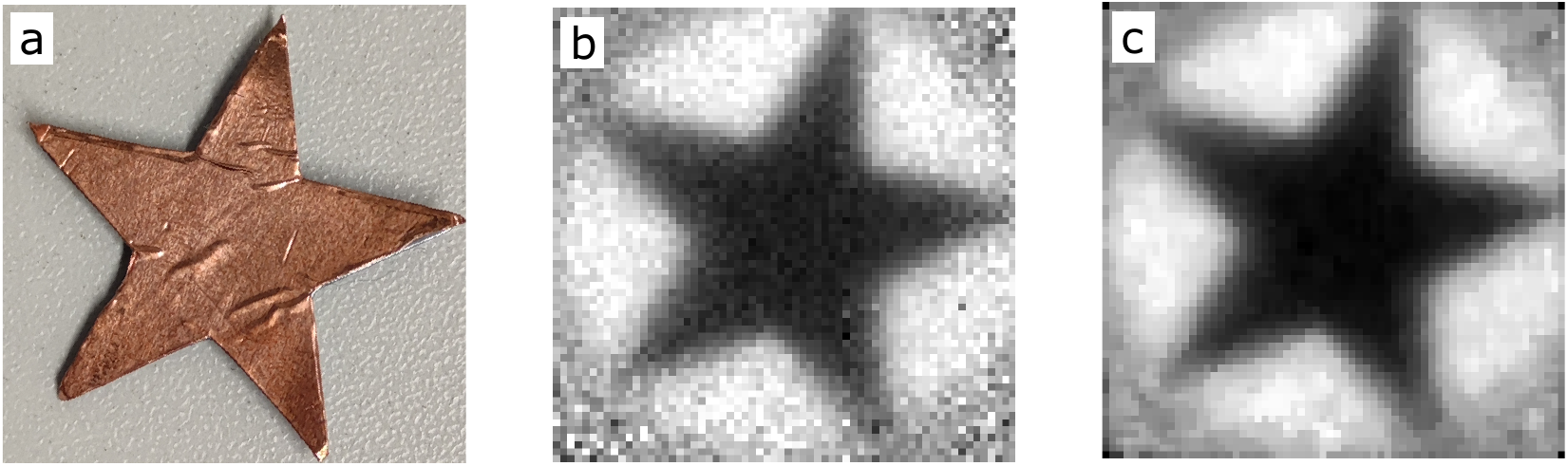}
\caption{(a) Photograph of a copper star. (b) 64x64 pixel, terahertz image of the star. (c) Filtered image with 3x3 mean filtering mask.}
\end{figure}

\newpage
\section{\bf 7. Spectral sensitivity}
\noindent{In our work we detected  signals with a broadband, direct-detection, incoherent detector (InSB bolometer) with spectral sensitivity in the range 0.06 - 1.5\,THz, but without any spectral resolution. Here we briefly show that the main contrast in the images comes from the radiation below 1\,THz, which is consistent with the dropping emissivity modulation at frequencies above 1\,THz  (see Fig.~1c of the main paper). We use 2 circular low pass THz filters with cutoff frequencies 1 and 2\,THz (see photograph Fig.~13a) and we place them right in front of the emissivity modulator - by plotting normalised images we can compare the relative amount of power passed through both the filters - see Fig.~13b. and c for the raw and filtered images. In Fig. 13d we plot 1D cuts of the signal on the diagonal, crossing through both the filters - we can see how the signal within the circular field of view is highest in the central part (~19\,mm), where there is a small gap between the filters. The signal coming from the area under the 1\,THz filter is about 10\% lower compare to the area below the 2\,THz filter.\\}

\begin{figure}[ht!]
\centering
\includegraphics[width=1\linewidth]{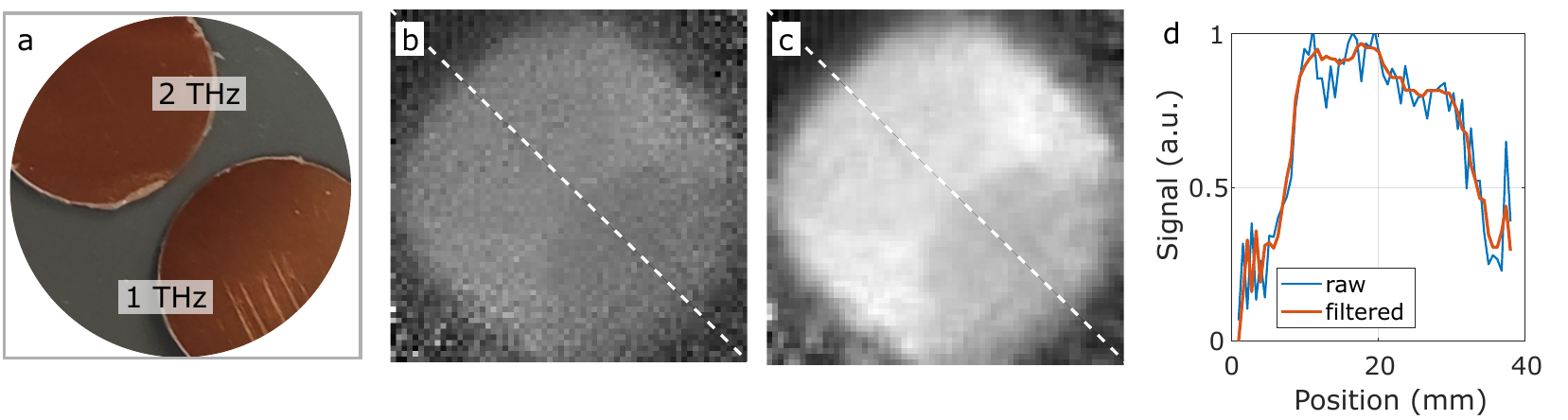}
\caption{(a) Photograph of the 1 and 2\,THz low-pass frequency filters inside the field of view. (b) Raw, unfiltered, normalised image of the filters and image obtained after mean filtering with 3x3 masks (c). (d) A cut through the field of view (dashed lines in b and c).}
\end{figure}

\section{\bf 8. Artefacts  from the specular reflection of the detector on the wafer}
\noindent{When the wafer is placed perpendicular to the optical axis (see Fig.~1), the detector itself is specularly reflected from the wafer and the signal becomes additionally patterned by the distribution of the low internal temperature in the detector's cryostat. This is due to the fact that the detector lies in the conjugate plane to the wafer. \\}

We became more sensitive to the emissivity/reflectance change of the wafer compared to the case described in Fig.1, where the detector sees a room temperature absorber at specular reflection. This could also be considered a technique to increase the contrast in the images.

This is the scenario under which a room temperature wafer can generate a measurable signal, unlike as described in the discussion part of the main paper. Fig.~14a shows an image acquired with a room temperature wafer normal to the optical axis and Fig.~14b an image corresponding to the tilted wafer. 

\begin{figure}[ht!]
\centering
\includegraphics[width=0.6\linewidth]{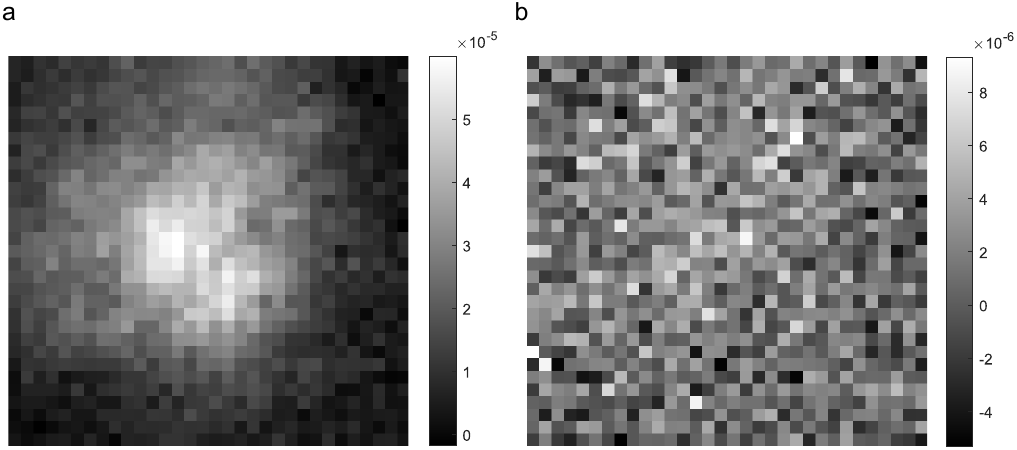}
\caption{(a) An artefact originating from the specular detector reflection with room temperature wafer normal to the optical axis. The low temperature distribution of the cryostat is projected onto the wafer and is consequently imaged by the detector. (b) With a small tilt of 15\,$^\circ$ of the wafer with respect to the optical axis. In this case the specular reflection is redirected to a room temperature absorber according to Fig.1.}
\end{figure}

\section{\bf 9. Estimating sensitivity of the detector}

\noindent{Here, we show results of indicative measurements of detector's responsivity $R$ and its noise spectrum density $V_\mathrm{n}$. Knowledge of these two parameters can then be used to estimate the noise equivalent power of the detector NEP, a measure of its sensitivity. \\}

The sensitivity of a detector is assessed by its NEP defined as the input power level that results in a signal-to-noise ratio of 1 at the end of an integration period $\Delta t_\mathrm{int}$

\begin{equation}
	\frac{S}{N} = \frac{P_\mathrm{in}}{NEP} \sqrt{2\cdot \Delta t_\mathrm{int}}
    \label{eq:NEP}
\end{equation}

The NEP can be evaluated by the following equation if the spectral noise density $V_\mathrm{n}$ and responsivity $R$ of the detector are known \cite{Rieke2021}:
\begin{equation}
  	NEP = \frac{V_\mathrm{n}}{R}.
\end{equation}

We measure the spectral noise density $V_\mathrm{n}$ at the output of the low noise amplifier with 1000x amplification. We perform the measurement in the time domain by sampling the output noise of the detector with sampling frequency Fs = 500 kHz over a time window of 1\,s (500\,k samples acquired). Upon Fourier transform of the time domain noise signal, we directly obtain the spectral noise density in V/Hz$^{1/2}$ as the frequency step in the frequency domain is exactly 1\,Hz. To sample to noise signal we use NI DAQ card  PCIe-6341 with 16 bit resolution and a minimum voltage range of $\pm$ 100 mV resulting in the theoretical rounding error / noise of $1/2 \cdot 200 \mathrm{mV} / 2^{16}$ = 1.55\,$\upmu$V. The internal RMS noise of the DAQ was measured to be 14\,$\upmu$V$_\mathrm{rms}$ (9\,$\upmu$V$_\mathrm{rms}$ according to the datasheet). The measured value corresponds to noise spectral density of V$_\mathrm{rms} / \sqrt{f_c}$ = 14$\upmu$V / $\sqrt{250000}$ = 0.89\,nV/Hz$^{1/2}$, $f_c$ being the cutoff frequency.
The measured noise density thus includes all noise contribution from the detector, its coupling optics, electronic noise of the amplifier and photon noise of the incident radiation (negligible) as the detector is directed toward a room temperature absorber (background noise).

Fig.~\ref{fig:noise} shows the spectral noise density at the output of the amplifier measured in the band 0-250\,kHz. Taking into consideration the gain of the amplifier (G = 1000)  we can calculate the approximate amplifier's input noise density (output noise density of the bolometer). The average noise density in the postdetection frequency band 0-250\,kHz is 9.2\,nV/Hz$^{1/2}$ at the output of the bolometer / input of the amplifier.

\begin{figure}[ht!]
\centering
\includegraphics[width=0.45\linewidth]{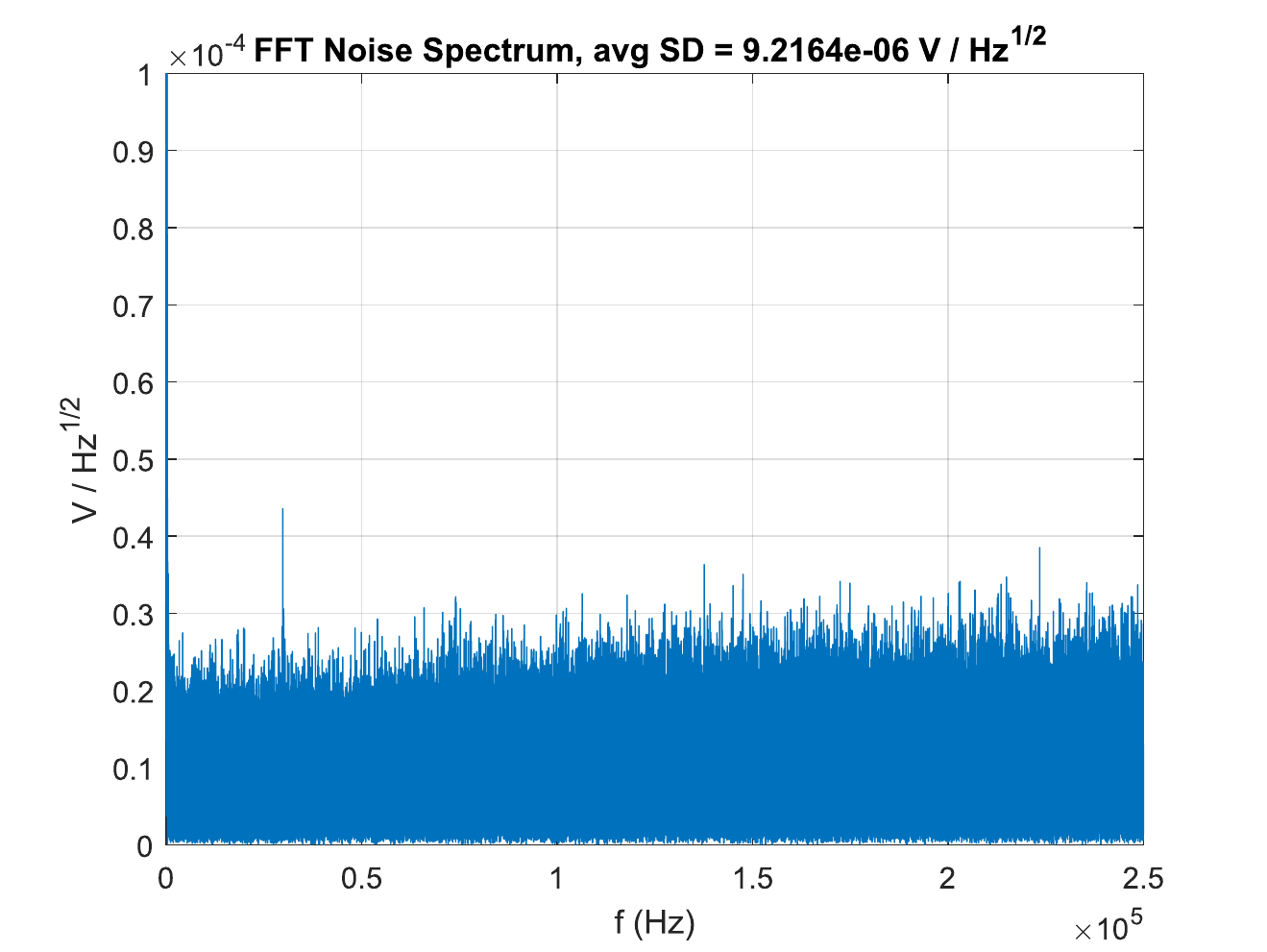}
\caption{Noise spectral density at the output of the low-noise amplifier.}
\label{fig:noise}
\end{figure}

In order to estimate the responsivity $R$ of the detector we use a commercial, coherent source based on a Gunn diode, Linwave Technology LW22-797599 generating a continuous wave radiation at $\nu_\mathrm{Gunn}$ = 107\,GHz with output power of 15\,mW. The Gunn diode is followed by a frequency trippler combined with a diagonal horn antenna from VDI (WR2.8X3HP) with 3.8\,\% conversion efficiency at the output frequency of 321\,GHz. The theoretical power in the beam is thus 570\,$\upmu$W. We calibrate the source using an absolute THz power meter Ophir 3-A-P-THz and measure the total power of 430\,$\upmu$W. The gain of the diagonal horn is 25\,dBi. As the output power of our THz source cannot be altered, we instead place the source on a linear translation stage and vary the distance between the source and the detector $d$ (i.e. aperture of the Winston cone inside of the cryostat) between 26.8 and 51.8\,cm. As the detector is only sensitive to change in the signal, the radiation from the source is chopped by a chopper with absorber covered blades to avoid any stray reflections from the conductive blades. Fig.~\ref{fig:responsivity}a plots the detected voltage as a function of the distance. A free-space path loss fit to the data is provided as well. We can see the slope of the experimental data to be smaller compared to the theoretical fit. Non-linearity of the detector is the most probable cause for this behaviour.

In Fig.~\ref{fig:responsivity}b we evaluate the detected, 1000x amplified voltage as a function of theoretical input power level corresponding to the distance between the source and the detector, assuming the gain of the source antenna is known $G_\mathrm{Tx}$=25\,dBi, the power transmitted as well $P_\mathrm{Tx}$ =  430$\upmu$W. The free-space path loss is given by $\left[\lambda/(4 \pi  d)\right]^2$ and the gain of the Winston cone collection optics coupled to the detector is calculated from  $G_\mathrm{Rx}=4\pi A_\mathrm{Rx}/\lambda^2$, where $A_\mathrm{Rx}$ correspond to the physical area of the aperture. The input power available to the detector $P_\mathrm{Rx}$ (i.e. $P_\mathrm{in}$ in eq. \ref{eq:NEP}), plotted on the horizontal axis of Fig.~\ref{fig:responsivity}b can be calculated from the Friis transmission equation as:

\begin{equation}
    P_\mathrm{Rx} =  P_\mathrm{Tx} \cdot G_\mathrm{Tx} \cdot G_\mathrm{Rx} \cdot \left[\lambda/(4\pi d)\right]^2.
\end{equation}

\begin{figure}[ht!]
\centering
\includegraphics[width=1\linewidth]{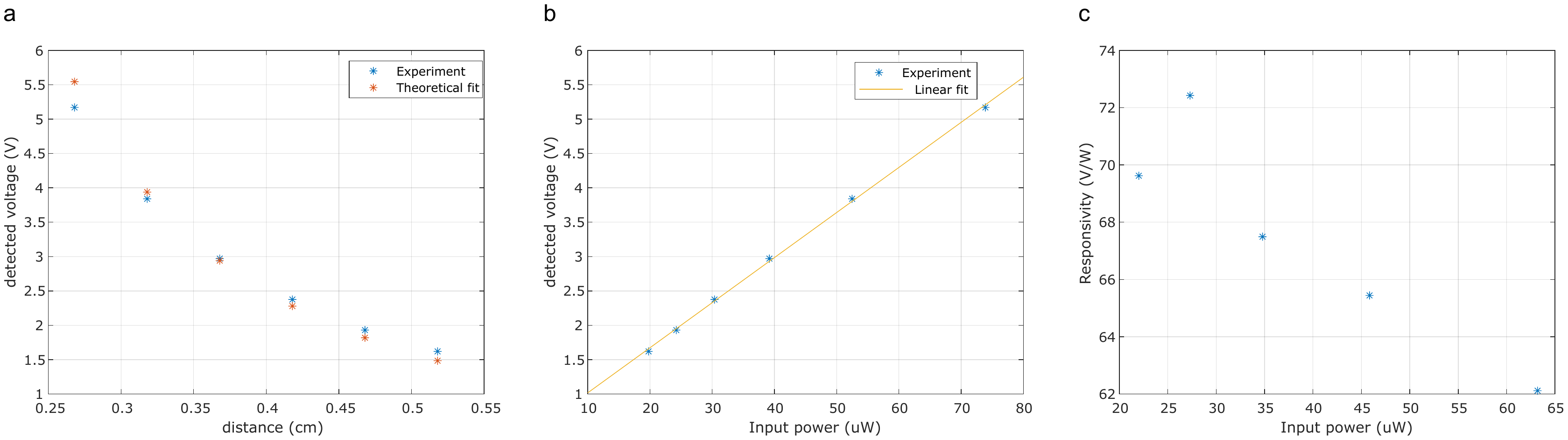}
\caption{(a) Measured voltage level as a function of the distance between the apertures of the source and detector. (b) Measured voltage as a function of predicted input power level. (c) Sensitivity of the detector as a function of the predicted power level for 100\,\% aperture efficiency of the detector}
\label{fig:responsivity}
\end{figure}

Fig.~\ref{fig:responsivity}c evaluates the final responsivity $R$ as a function of the predicted input power as $R = \mathrm{d}V_\mathrm{out} / \mathrm{d}P_\mathrm{Rx}$, where $\mathrm{d}V_\mathrm{out}$ is the change in detected voltage at the input of the amplifier (i.e. dividing the value of the voltage in Fig.~\ref{fig:responsivity}b by a factor of 1000) as a function of input power change $\mathrm{d}P_\mathrm{Rx}$. All the coupling losses in the detector's system - Winston cone coupling, dielectric losses in the windows are accounted for in this value. 

\textbf{The NEP can be evaluated from the measured noise spectral density $V_\mathrm{n}$=9.2\,nV/Hz$^{1/2}$ and responsivity estimate $R\approx 70$\,V/W as NEP = $V_\mathrm{n}/R = (9.2$\,nV/Hz$^{1/2})$/(70\,V/W) = 130\,pW/Hz$^{1/2}$.}

We note that the Winston cone (i.e. the detecting antenna) is not a standard single mode antenna but rather a multi-mode collection optical element \cite{Smestad1990} and the standard near-field range limit on such measurement, in which the distance between the source and detector antennas must be larger than $2D^2/\lambda$, where $D = 25$\,mm is the diameter of the Winston cone, does not have to be satisfied. Instead the numerical aperture of the Winston cone must be larger than the numerical aperture of the measurement geometry. In our case, the half-width at half-maximum (HWHM) angle of the Winston cone is 14\,$^\circ$ corresponding to the NA$_\mathrm{Winston} = 0.24$ and the HWHM angle corresponding to the line drawn from the position of the Tx antenna to the edge of the Winston cone is 2.6\,$^\circ$ for the shortest distance $d$=26.8\,cm which corresponds to the NA$_\mathrm{exp} = 0.045$. We see that the requirement on the NA is met with sufficient margin.\\

\noindent {\bf Photon noise contribution}\\
In our experiments, the photon noise is negligible with approximate NEP \cite{Rieke2021} contribution of 
\begin{equation}
  	NEP_\mathrm{ph} = \frac{h c}{\lambda} \left(\frac{2\phi}{\eta}\right)^{1/2},
\end{equation}
where $ h, c $ represent the Planck’s constant and speed of light, $\lambda$ is the central wavelength (0.38\,mm at 0.78\,THz), $\phi$ is the number of photons per second incident on the detector and $\eta$ is the quantum efficiency, which is relatively high for bolometers (90-100 \% \cite{Rieke2021}). The above given equation is an approximation as it disregards the Bose-Einstein bunching nature of photons. 
The photon flux can be evaluated from the average power incident on the detector which, under the Rayleigh-Jeans approximation to the Planck’s law can be calculated as:
\begin{equation}
 P = \frac{2}{3c^2} \cdot A \cdot \Omega \cdot k_\mathrm{B} \cdot T \cdot [\nu_2^3 - \nu_1^3], 
\end{equation}
where $A$ is the projected area corresponding to the object / field of view, $\Omega = a/r^2$ is the solid angle corresponding to the size of the detector (area $a$) at a distance $r$ away from the object, $k_\mathrm{B} = 1.38\cdot 10^{-23}$ J/K is the Boltzman’s constant, $T$ is the thermodynamic temperature of the object and $\nu_2$ and $\nu_1$ are the upper and lower frequency limit of the detector. In our experiment, it follows $A$ = 0.1963\,m$^2$ since the distance from the detector to the room temperature (T = 293 K) absorber is 1\,m and the detector’s numerical aperture is NA = 0.2426 corresponding to the maximum angular range of $\pm 14^{\circ}$. Diameter of the detector’s aperture (Winston cone) is 25\,mm with the detector's solid angle $\Omega = a/r^2 = 4.91 \cdot 10^{-4}$ sr. Upper and lower frequency limits are $\nu_1$ = 0.06 THz and $\nu_2$=1.5 THz. For this set of parameters and by neglecting any coupling loss in the optics (e.g. windows) and a flat frequency response we arrive at the input power level of 9.74 $\upmu W$. Correspondingly, the photon flux is given as $\phi = E_\mathrm{tot}/E_\mathrm{ph}$, where $E_\mathrm{tot}$ is the total energy given by the power $P$ and the integration period $\Delta t_\mathrm{int}$ and $E_\mathrm{ph} = h\, \nu_0 = 5.17\cdot 10^{-22}$\,J is the energy per photon at the center frequency $\nu_0 = 0.78$\,THz. For 1\,s integration time $\phi = 1.88\cdot10^{16}$ photons per second correspond to the photon noise equivalent power of $NEP_\mathrm{ph}$ =0.1\,pW/Hz$^{1/2}$.\\

\end{document}